# Stability Diagram of a Few-Electron Triple Dot

L. Gaudreau

Institute For Microstructural Sciences, NRC, Ottawa, Canada K1A 0R6

Régroupement Québécois sur les Matériaux de Pointe,

Université de Sherbrooke, Quebec, Canada J1K 2R1

S. A. Studenikin, A. Sachrajda, P. Zawadzki, A. Kam, J. Lapointe, M. Korkusinski and P. Hawrylak

Institute For Microstructural Sciences, NRC, Ottawa, Canada K1A 0R6

Individual and coupled quantum dots containing one or two electrons have been realized and are regarded as components for future quantum information circuits. In this work we map out experimentally the stability diagram of the few electron triple dot system, the electron configuration map as a function of the external tuning parameters, and reveal experimentally for the first time the existence of quadruple points, a signature of the three dots being in resonance. In the vicinity of these quadruple points we observe a duplication of charge transfer transitions related to charge and spin reconfigurations triggered by changes in the total electron occupation number. The experimental results are largely reproduced by equivalent circuit analysis and Hubbard models. Our results are relevant for future quantum mechanical engineering applications within both quantum information and quantum cellular automata (QCA) architectures.

A comparison between single quantum dots and real atoms confirms both analogous and dissimilar properties. Atomic-like shell structure and Hunds rules govern both systems[1,2]. The very different energy scale of the artificial atom, however, manifests itself in novel interaction phenomena which have no analogue in real atoms, such as singlet triplet transitions and spin texture arrangements of electrons[3,4]. Recently, with the realization of electrostatic few-electron quantum dots[5] and their combination with non-invasive charge detection technology[6], tuneable coupled few-electron quantum dots (i.e. artificial diatomic molecules) have been investigated. The tunability of these devices makes them promising candidates for future quantum information applications[5,7,8,9,10,11] as well as for fundamental studies of quantum molecular effects, and for exploiting nanospintronic functionalities[12]. There has been growing theoretical interest in the next level of complexity, the few-electron triple dot or triatom. In addition to applications in the field of quantum information as entanglers[9] or coded qubits[8,10,11], rectification and ratchet functionalities have been predicted[13,14]. We present here the first experimental results from a few electron triple dot potential formed by surface gates.

The starting point for any experimental investigation of a new complex quantum dot system is the stability diagram, the configuration map. In this paper we study for the first time the stability diagram of a few electron triple dot potential. We identify quadruple points where all three quantum dots are simultaneously in resonance. We find these are characterised by internal charge and spin rearrangements. A dramatic manifestation of these rearrangements are *duplications* or *clones* of charge transfer boundaries (i.e. when electron configurations switch at fixed total electron number). We concentrate on the most fundamental quadruple points in which one, two, or three electrons are shared between the three dots. While these new effects will need to be taken into account when designing either charge or spin qubit implementation schemes, they also provide new opportunities for processing and moving quantum information around a circuit. Additionally, this phenomenon may be considered a basic demonstration of a QCA effect in a few electron regime and adaptations may form the basis of future QCA architectures[15,16].

Ciorga et al.[17] first demonstrated that a single electron could be isolated from a two dimensional electron gas in an AlGaAs/GaAs heterostructure by purely electrostatic means. By slightly adapting the gate layout (i.e. in effect slicing the few electron quantum dot potential in two) two coupled few electron quantum dots were soon after realised[12,18,19,20,21,22,23,24]. In fact modelling the confinement potential of a few electron double dot layout reveals[25] that one should, by applying suitable gate voltages, be able to create a few electron triple dot (where the dots are aligned in series). On three separate cooldowns of our device, however, measurements confirmed that while we were able to create a triple dot few electron potential, remarkably the three dots arranged themselves in a ring. In this paper we take advantage of this to study this novel quantum dot circuit for the first time. A schematic of the triple dot potential is shown in the inset of figure 1. All three dots A, B and C are coupled to each other. A quantum point contact (QPC) charge detector is used to probe changes in the triatom charge configurations[6]. The current through the QPC ($I_{QPC}$) is sensitive to any change in the charge configuration. The derivative of $\partial I_{QPC}/\partial V_{4B}$ is measured using standard low frequency AC techniques.

The stability diagram for a quantum dot circuit consisting of N coupled dots is an N-dimensional entity. Due to cross capacitances any 2D slice through the stability diagram should reveal N sets of parallel Coulomb blockade lines (each set reflecting when the electrochemical potential of the leads and a particular dot are matched.) To observe unique triple dot features, it is necessary to achieve the very particular slices in which lines from all three sets cross at a single location. These are called the quadruple points where four electron configurations are degenerate.

The above predicted behaviour is confirmed in the stability diagram in figure 1. The three sets of parallel lines marked by red, blue and green circles refer to changing the number of electrons in A, B and C respectively. The three ovals indicate triple points associated with the three possible two-dot combinations. The first successful realisation of the fundamental configurations (0,0,0) and (1,1,1) can be seen. The (1,1,1) configuration is a necessary condition to create the Greenberger-Horne-Zeilinger maximally entangled states, using which quantum algorithms, such as quantum teleportation can be implemented[26].

To realise quadruple points experimentally, two triple points need to be merged. This is achieved by tuning voltages $V_{3T}$ and $V_{3B}$ and is shown in detail in figures 2(a) to 2(d). The blue and pink circles in fig.2 marks triple points between A and B in which configurations (0,1,0) and (0,0,0) are degenerate with (1,0,0) and (0,0,1) respectively. In figure 2(a) these points are still far apart. A charge transfer line is indicated by dotted lines for clarity ((1,0,0) to (0,1,0). In figure 2(b) the two triple points are adjusted closer together. On close inspection, a new feature is emerging close to the C=1 line (the line corresponding to the addition of the first electron into C) at the division between (0,1,0) to (0,1,1) states. In figure 2(c) this extra feature has evolved into a new additional charge transfer line, parallel to the one between (1,0,0) and (0,1,0). Based on the models presented below the configurations on either side of this 'clone' line are identified as (1,0,1) and (0,1,1). In figure 2(d) we observe that the original charge transfer line between (1,0,0) and (0,1,0) disappears at the C=1 line.

We first evoke an equivalent circuit model[27], in which the electrostatic energy of the system is given by $E_{triatom} = \sum_{i=1}^{3}(\frac{1}{2}U_i N_i^2) + \sum_{i<j,j=2}^{3} N_i N_j U_{i,j} + f(V_{1B}, V_{5B})$, where $N_i$ is the number of electrons in dot $i$, $U_i$ is its charging energy, $U_{i,j}$ are the capacitive coupling energy between dots $i$ and $j$, and $f$ is the electrostatic energy due to the gate voltage induced charge. The ratios of charging and capacitive coupling energies can be estimated directly from the data in Fig.2. Specifically, $U_A$=3.1 meV, $U_B$=2.8 meV, $U_C$=2.28 meV, $U_{AB}$=0.86 meV, $U_{AC}$=0.28 meV, $U_{BC}$=0.57 meV. To include quantum effects we also develop a Hubbard model. This model includes tunnelling and the Pauli exclusion principle. With $c_{i\sigma}^{+}$ ($c_{i\sigma}$) operators describing the creation (annihilation) of an electron at the lowest energy level $E_i$ of the $i^{th}$ dot ($i$ = A,B,C), the Hubbard Hamiltonian can be written as $H = \sum_{ij\sigma}(E_i\delta_{ij} + t_{ij})c_{i\sigma}^{+}c_{j\sigma} + \sum_i U_i n_{\uparrow i} n_{\downarrow i} + \sum_{i<j} V_{ij} n_i n_j$, where $n_{\uparrow i} = c_{i\uparrow}^{+}c_{i\uparrow}$ and $n_i = n_{\uparrow i} + n_{\downarrow i}$ is the spin and charge density in dot "$i$", and $t_{ij}$ is the interdot tunneling ($t_{AB}$=70 μeV , $t_{BC}$=10μeV. $t_{AC}$=10μeV ). The three terms describe the energy of each dot, the charging energy and the interaction energy between electrons occupying different dots. To make contact with experiment, we assume that the energies $E_i$ are linear functions of gate voltages; other Hubbard parameters are extracted from experiment. The ground-state energy $E_0$ is obtained by diagonalizing the Hubbard Hamiltonian. These energies are used to calculate the chemical potential $\mu(N)$. When $\mu(N)$ equals the chemical potential of the leads, the N+1$^{st}$ electron is added and the stability diagram is established. From the corresponding ground-state wave function we determine the occupation $n_A$, $n_B$, $n_C$ of each dot and compute the QPC voltage $V_{QPC} \sim n_A/R_A + n_B/R_B + n_C/R_C$, with $R_A$, $R_B$, $R_C$ being respectively the distances of the A, B, and C dots from the QPC. Figures 3(b) and 3(c), which should be compared with figure 2(c), show the results from the two models. The cloning of the charge transfer line is confirmed by both models. Quantum effects included in the Hubbard model lead to a smearing and curvature of the charge addition and transfer lines.

The origin of the duplicate charge transfer line is revealed as a charge re-arrangement effect, reminiscent of QCA. This is shown in more detail in figure 4(a). Initially, a single electron resides in A. As gate 5B is made less negative it becomes energetically favourable for the electron to transfer to B. As the gate sweep continues, the system enters the two-electron regime involving a more complex, two-step process. When the second electron enters C, the electron in B reverts back to A. This is a consequence of the difference in interdot coupling between A and C, and B and C. Finally, as the gate is swept further, the electron in A is once again transferred into B. This results in a line parallel to the first charge transfer line since they both involve the transfer of an electron from A to B, differing only by the presence of an electron in C. The disappearance of the original charge transfer line at the C=1 line is now also clear. The configuration on crossing the C=1 line is uniquely the (1,0,1) state for both the initial (1,0,0) and (0,1,0) configurations. The models also explains other observations such as the modification of the slope of the C=1 line between the two quadruple points marked $\alpha$ and $\beta$ and the behaviour of the slopes of two neighbouring charge transfer lines marked by $\gamma$ and $\delta$ in figure 2(a). Consider, for example, the latter effect. These are the lines obtained when transferring an electron into an empty dot C from A or B. Comparing figure 2(a) and 2(d) the $\gamma$ and $\delta$ slopes are reversed. But in 2(c) they have exactly the same slope. In 2(a), $\gamma$ ($\delta$) refers to a transfer of an electron from A (B) to C. But in 2(c), as a result of the charge reconfiguration effect, both lines involve a transfer of an electron from B into C resulting in similar slopes. Certain features remain to be fully understood.

Both models for example, predict that the clone charge transfer will *always* be parallel to its partner while the experiment reveals that initially (i.e. when it is close to the C=1 line) its slope is close to the modified C line slope and not the original charge transfer line.

We are able to successfully realise many quadruple points experimentally by gate adjustments. At the most fundamental quadruple point, $\alpha$, the 'vacuum' state (0,0,0) is resonant with the three single electron configurations (1,0,0), (0,1,0) and (0,0,1). At quadruple point $\beta$ the (1,1,1) configuration is resonant with the three double occupied states (1,1,0), (0,1,1) and (1,0,1). At other quadruple points simultaneous charge rearrangements are expected to be accompanied by spin transitions. In figure 4(b), for example, the charge rearrangement involves the transition from (0,2,0) to (1,1,1) which, at low magnetic fields, one would expect to be accompanied by a spin flip event[24,3] in dot B. The schematics indicate the distribution of electrons amongst the dots.

In conclusion, the stability diagram of a few-electron triple dot has been measured for the first time. The results reveal a new playground for fundamental physics and open up new opportunities for fundamental studies of complex entangled quantum states. The stability diagram is found to be a complex object containing charge and spin rearrangements in the vicinity of quadruple points. These rearrangements may lead to applications in QCA circuits as well as in quantum information implementation schemes.

A.S. acknowledges financial support from the Natural Sciences and Engineering Research Council of Canada and A.S., P.H. and A. Kam acknowledge support from the Canadian Institute for Advanced Research. The authors thank K. Le Hur and M. Pioro-Ladriere for discussions.

Fig 1.

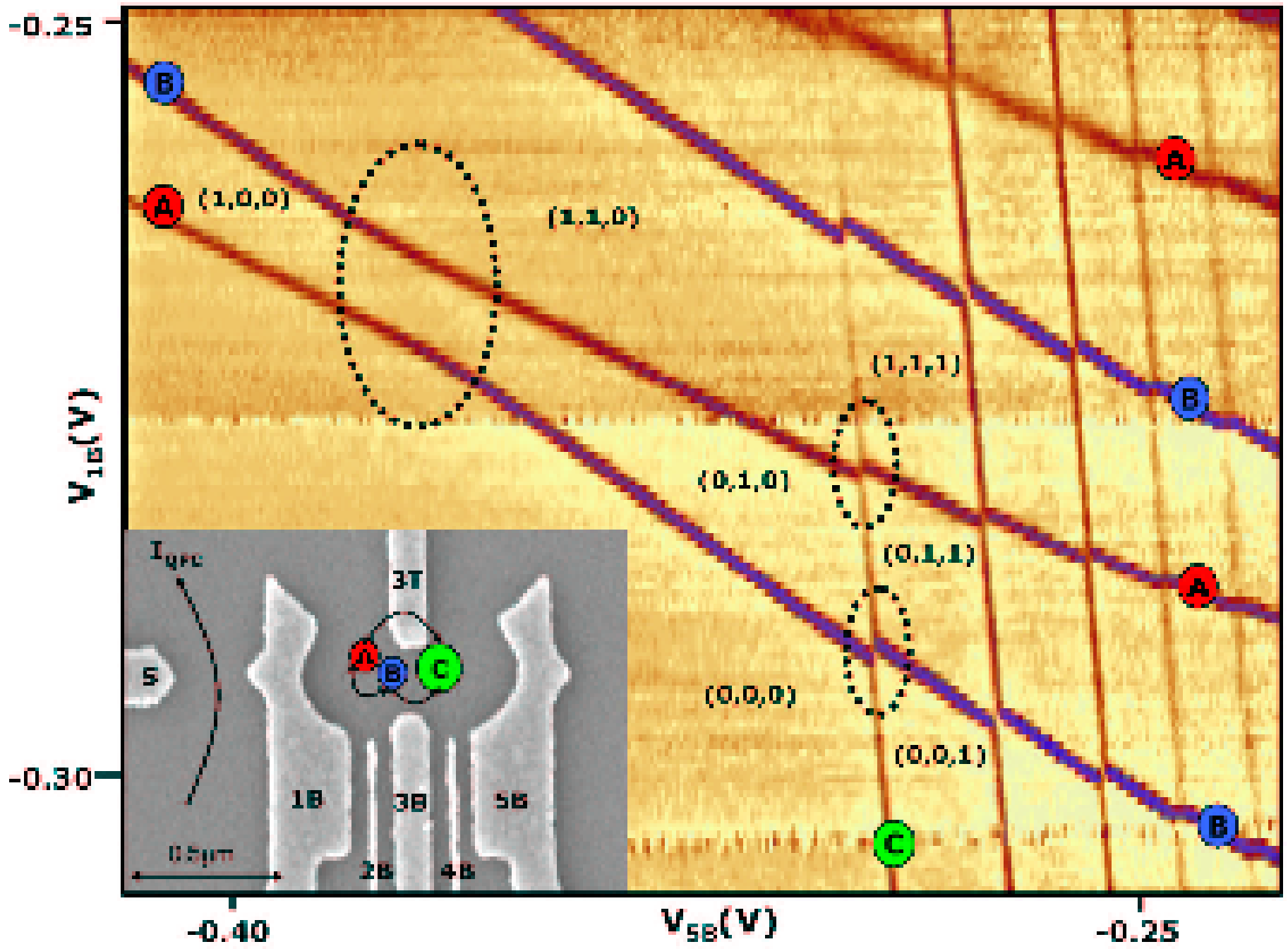

**Figure 1. (Color online) Stability diagram of the triatom. Three different sets of parallel lines correspond to the addition spectrum of dot: A (red), B (blue) and C (green). Dashed ovals correspond to three different triple points, where chemical potentials of two dots and the leads coincide. Inset, Scanning electron micrograph of a device, schematically indicating the position of the three dots.**

Fig.2

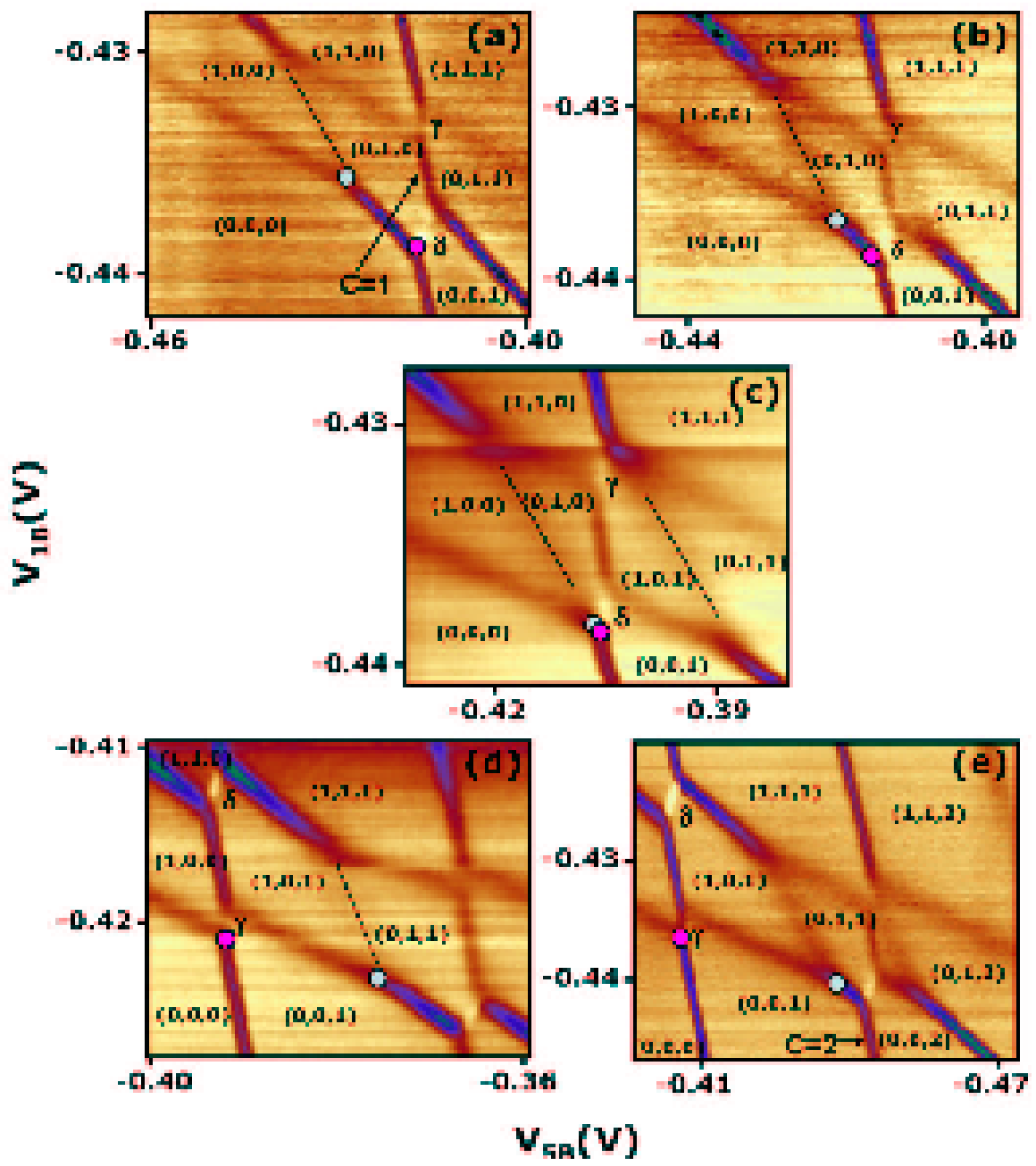

**Figure 2. (Color online) Evolution in the (0,0,0) to (1,1,1) region as two triple points are gradually merged. Dashed lines have been drawn over inter-dot charge transfer lines for clarity. a, Triple points where dots A and B are in resonance (light blue circle) and dots B and C are in resonance (pink circle). b, The two triple points are closer together and a new structure starts to appear on the right side of the C=1 line. c, The two triple points coincide, creating a quadruple point where the three dots are in resonance. We observe the appearance of a new line on the right side of the C=1 line. d, The triple points move apart again, the original charge transfer line has disappeared.**

Fig 3.

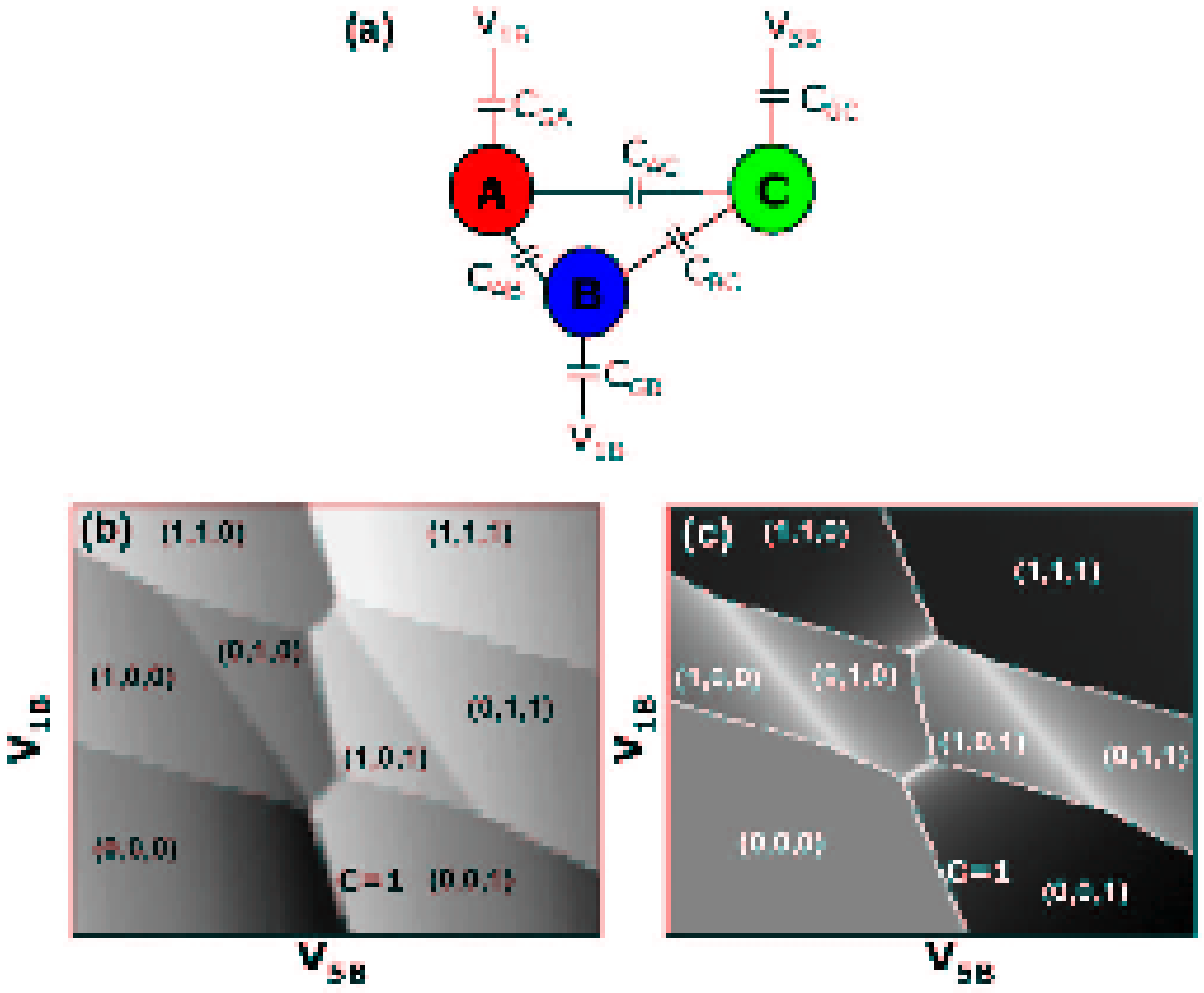

**Figure 3. (Color online) Theoretical models for the triple dot. a, Schematics of the circuit model. b, Theoretical stability diagram using the equivalent circuit model c, Hubbard model stability diagram.**

Fig. 4

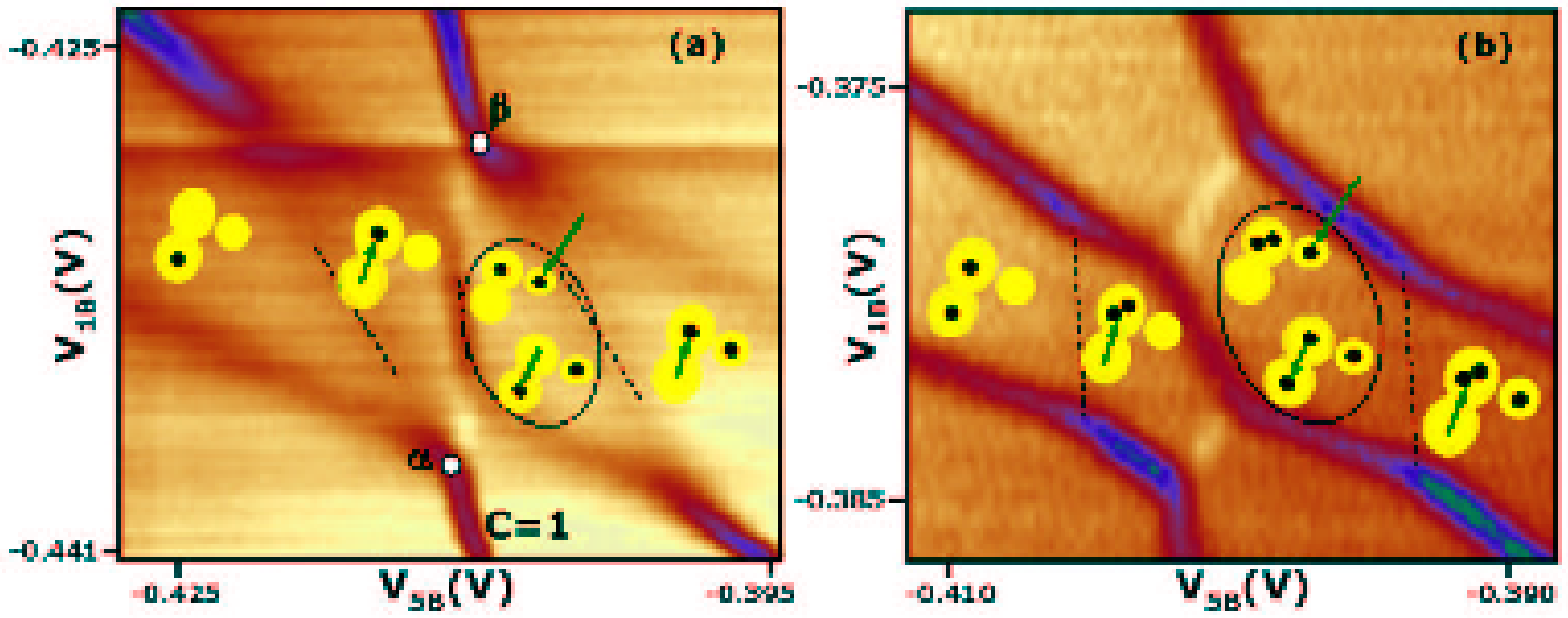

**Figure 4. (Color online) Schematic evolution of the triatom configuration in the vicinity of two different quadruple points as gate 5B is made less negative. a, Starting with one electron in A  (1,0,0), the electron is transferred to B (0,1,0), the addition of an extra electron to C triggers an electronic rearrangement in the other dots and the first electron is transferred from B to A (1,0,1), finally the electron is retransferred from A to B (0,1,1) . b, Starting with one electron  in A and 1 electron in B (1,1,0), the electron from A is transferred to B (0,2,0), the addition of an extra electron in C creates again a charge rearrangement (1,1,1), but this time involving a spin rearrangement at low magnetic fields. Finally an electron is transferred from A to B again (0,2,1).**

---